\documentclass[aoas,preprint]{imsart}

\RequirePackage[OT1]{fontenc}
\RequirePackage{amsthm,amsmath}
\RequirePackage{natbib}
\RequirePackage[colorlinks,citecolor=blue,urlcolor=blue]{hyperref}
\usepackage{graphicx}
\usepackage{bbm}

\startlocaldefs
\numberwithin{equation}{section}
\theoremstyle{plain}

\endlocaldefs

\newcommand{\bftheta}{{\boldsymbol \theta}}

\newcommand{\bfs}{{\bf s}}
\newcommand{\bfSigma}{{\bf \Sigma}}

\begin{document}

\begin{frontmatter}
\title{Review: Nonstationary Spatial Modeling, with Emphasis on Process Convolution and Covariate-Driven Approaches}
\runtitle{Review: Nonstationary Spatial Modeling}

\begin{aug}
\author{\fnms{Mark D.} \snm{Risser}\ead[label=e1]{mdrisser@lbl.gov}}

\runauthor{M. D. Risser}

\address{Climate \& Ecosystem Sciences Division\\
Lawrence Berkeley National Laboratory\\
1 Cyclotron Road\\
Berkeley, CA 94720\\
\printead{e1}}

\end{aug}

\vskip2ex

\begin{abstract}
In many environmental applications involving spatially-referenced data, limitations on the number and locations of observations motivate the need for practical and efficient models for spatial interpolation, or kriging. A key component of models for continuously-indexed spatial data is the covariance function, which is traditionally assumed to belong to a parametric class of stationary models. While convenient, the assumption of stationarity is rarely realistic; as a result, there is a rich literature on alternative methodologies which capture and model the nonstationarity present in most environmental processes. This review document provides a rigorous and concise description of the existing literature on nonstationary methods, paying particular attention to process convolution (also called kernel smoothing or moving average) approaches. A summary is also provided of more recent methods which leverage covariate information and yield both interpretational and computational benefits.

\textbf{Note:} the following is borrowed from Chapters 1 and 2 of the author's Ph.D. dissertation (\citealp{MRisserThesis}), joint with Catherine A. Calder.  
\end{abstract}


\end{frontmatter}

\section{Introduction: spatial statistical modeling}

Despite the rising popularity of spatio-temporal statistical modeling, there is still a strong need for flexible and computationally efficient spatial models appropriate for spatial prediction. For example, in the case of meteorological, agricultural, or geological data where fixed monitoring stations are used to collect observations of a spatial process, monitoring sites are not always located where information about the spatial process is desired. In such situations, it may be of interest to generate a ``filled-in'' prediction map of the spatial process based on a sparse, finite number of observations, as well as estimate the uncertainty in these predictions. 

The standard way to model a point-referenced, continuously-indexed spatial process is to specify that observations of the process are generated by a particular stochastic mechanism or stochastic process. A Gaussian process (GP) is an extremely popular choice for the stochastic process, due to the fact that all finite-dimensional distributions are known to be Gaussian and because the process is completely specified by a characterization of its first- and second-order properties. Furthermore, for a GP, the second-order properties can be easily specified by one of the widely used classes of valid spatial covariance functions. The spatial covariance function describes the degree and nature of spatial dependence (or covariance) present in a spatial process.  In fact, a hallmark principle of spatial statistics states that, in general, values of the spatial process which are close together are more likely to be similar (dependent), while values which are far apart are most likely unrelated (approximately independent). 

\subsection{Second-order properties of spatial processes}\label{2ndOrderProps}

Define $\{ Y(\bfs) : \bfs \in G\}$ to be a general spatial stochastic process of interest, where the spatial domain $G \subset \mathcal{R}^d, d \geq 1$. Furthermore, without loss of generality, assume that the process $Y(\cdot)$ is mean-zero, i.e., $E[ Y(\bfs) ] = 0$ for all $\bfs \in G$. Define $C(\cdot \>, \> \cdot)$ to be the spatial covariance function of $\{ Y(\bfs) : \bfs \in G\}$, such that 
\[
C(\bfs, \bfs') \equiv \text{Cov}[ Y(\bfs), Y(\bfs') ] = E[Y(\bfs) \cdot Y(\bfs')],
\]
for all $\bfs, \bfs' \in G$. The covariance function $C$ is always symmetric, i.e., $C(\bfs, \bfs') = C(\bfs', \bfs)$, and when $\bfs = \bfs'$, the covariance function defines the variance of the process
\[
C(\bfs, \bfs) = \text{Var}[Y(\bfs)].
\]
The covariance function must be a nonnegative definite function, meaning that
\begin{equation} \label{PD}
\sum_{i=1}^n \sum_{j=1}^n a_i \bar{a}_j C({\bf s}_i, {\bf s}_j) \geq 0,
\end{equation}
for any positive integer $n$, any set of locations $\{ {\bf s}_i : i = 1, \dots , n\} \in G$, and any set of complex numbers $\{a_i : i = 1, \dots, n\}$ ($\bar{a}_i$ denotes the complex conjugate of $a_i$). 

In order to learn about the covariance function from realizations of a spatial process, further assumptions are nearly always made regarding the properties of $C$. The most common is that of stationarity, meaning that some features of $C$ do not depend on the spatial location. More formally, a process $\{ Y(\bfs) : \bfs \in G\}$ is said to be second-order stationary (or weakly stationary) if the following two properties hold:
\begin{enumerate}
\item $E[Y(\bfs)] = E[Y(\bfs + {\bf h})] = c$ for some constant $c$, and 
\item $C(\bfs, \bfs + {\bf h}) = C({\bf 0, h})$ for some spatial lag ${\bf h} \in \mathcal{R}^d$. 
\end{enumerate}
For fixed ${\bf h}$, note that $C({\bf 0, h})$ is a constant which does not depend on $\bfs$. The covariance function for a spatial process which is second-order stationary can be written as $C(\bfs, \bfs') = C(\bfs - \bfs')$ and is often simply called a stationary covariance function. 
The first requirement is not restrictive since we have already specified $Y(\cdot)$ to be mean-zero, and non-constant mean behavior can be introduced into a different component of a statistical model. However, the second requirement is much more restrictive, as it is rarely reasonable to assume that the spatial dependence structure does not depend on spatial location.

A complete characterization of the class of valid covariance functions is given by Bochner's theorem (\citealp{BochnerBook}; \citealp{AdlerBook}), which states that a real-valued function defined on $\mathcal{R}^d$ is the covariance function of a stationary process if and only if it is even and nonnegative definite. Bochner's theorem is a powerful result, as it enables the construction of stationary processes by utilizing the existing literature on nonnegative definite functions.

Two special cases of second-order stationary processes are isotropic processes and anisotropic processes. An isotropic process has a covariance function which can be written in terms of the length of the spatial lag $||{\bf h}||$, or 
\begin{equation} \label{isotropicCov}
C(\bfs, \bfs + {\bf h}) = C(||{\bf h}||),
\end{equation}
where $|| \cdot ||$ represents the Euclidean norm in $\mathcal{R}^d$, i.e., $||{\bf x}|| = \sqrt{ \sum_{k = 1}^d x_k^2 }$. Isotropic processes are particularly restrictive, as not even directionality impacts the covariance function. Anisotropic processes are a slight generalization of isotropic processes, in that both distance and direction are incorporated into the covariance function by way of a linear transformation of the lag vector ${\bf h}$. That is, the covariance function can be written
\begin{equation} \label{anisotropicCov}
C(\bfs, \bfs + {\bf h}) = C(||{\bf A}^{-1/2}{\bf h}||),
\end{equation}
where ${\bf A}$ is a $d\times d$ positive definite matrix (often called the anisotropy matrix) which allows the range of dependence to be longer or shorter in particular directions. Intuitively, the isotropic covariance function (\ref{isotropicCov}) yields spherical correlation patterns, while the anisotropic covariance function (\ref{anisotropicCov}) yields ellipsoidal correlation patterns.

\subsection{Parametric models for stationary spatial covariance functions} \label{statCovModels}

Several parametric models for isotropic covariance functions are particularly popular in spatial statistical modeling. In a traditional framework, each of these depend on (at least) three parameters $\sigma^2$, $\tau^2$, and $\phi$, which represent the process variance, nugget, and range, respectively, all of which must be greater than zero. However, in more modern hierarchical modeling frameworks, it is often preferable to separate the nugget from the covariance function model and incorporate it in a different component of the model. Aside from the resulting hierarchical structure, this decision is made based on physical meaning, since in spatial modeling the nugget $\tau^2$ represents measurement error (although it also accounts for microscale variability that cannot be accounted for based on the resolution of the data). Thus, the nugget is often included in a model for observed data, rather than in a theoretical latent process model. In what follows, we will opt for the hierarchical model specification, which separates the nugget from the covariance function itself. Using this framework, the process variance (also called the partial sill) is $C({\bf 0}) = \sigma^2$ and represents the variability in the process. The range parameter $\phi$ does not directly represent the range of the covariance function, but instead determines how quickly the covariance function decays to zero. Smaller values of $\phi$ correspond to a faster decay, while larger values of $\phi$ correspond to slower decay. A summary of commonly used parametric isotropic covariance functions is available in chapter 2 of \cite{BanerjeeBook}. 

One of the most popular parametric models is the Mat\'ern covariance function (e.g., \citealp{SteinBook}), which depends on an additional smoothness parameter $\kappa > 0$. The Mat\'ern covariance function is
\begin{equation} \label{matern}
\mathcal{M}_\kappa(h) = \sigma^2 \frac{1}{\Gamma(\kappa)2^{\kappa - 1}} \left(  h/\phi \right)^\kappa K_{\kappa} \left( h/\phi  \right), \hskip3ex h \geq 0,
\end{equation}
where $K_\kappa(\cdot)$ denotes the modified Bessel function of the third kind of order $\kappa$. Two special cases of the Mat\'ern covariance function are when $\kappa = 0.5$, which results in the exponential covariance function
\[
\mathcal{M}_{0.5}(h) = \sigma^2 \> \exp \left\{ -h/\phi \right\}, \hskip3ex h \geq 0,
\]
and letting $\kappa \rightarrow \infty$, which results in the Gaussian covariance function
\[
\mathcal{M}_{\infty}(h) = \sigma^2 \> \exp \left\{ -(h/\phi)^2 \right\}, \hskip3ex h \geq 0.
\]

\section{Deformation, basis function, and Markov random field methods}
\label{ch2LR}

While assumptions of second-order stationarity for a Gaussian process are both convenient and widely made, these assumptions are almost never appropriate in real-world applications. Instead, a spatial process will almost always display some sort of nonstationarity, in which features of the process vary over space. In some cases the stationary and nonstationary components of a process can be separated, such that the first-order properties are nonstationary (spatially-varying) and a stationary covariance function is used, but in most cases even this assumption does not truly reflect the expected behavior of a spatial process. 

As a result, there is a rich literature on alternative methodologies for modeling the second-order nonstationarity present in most problems. Three approaches for introducing nonstationarity into a covariance function model are deformation methods, basis function expansions,  and Markov random field methods using stochastic partial differential equations (SPDEs); see Section \ref{kernConv} for a final approach. In what follows, the spatial domain of interest $G$ will be of dimension $d$, i.e., $G \subset \mathcal{R}^d$; without loss of generality it will be assumed that the spatial process is mean-zero.

\subsection{Deformation methods}

{\sloppypar
One of the earliest methods for introducing nonstationarity into a spatial model is known as the deformation method, due to \cite{sampGutt}. The fundamental idea of isotropic models is that the covariance between observation locations is a function of Euclidean distance and so, intuitively, the deformation method obtains a nonstationary covariance structure by rescaling interpoint distances in a systematic way over $G$. More formally, deformation involves transforming the geographic region of interest ($G$) to a different ``deformed'' space (say, $D$) wherein isotropy holds. The transformation $\xi: G \rightarrow D$ is ideally one-to-one and, in general, a nonlinear mapping. Formally, the covariance of the spatial process $Y(\cdot)$ between two locations ${\bf s}, {\bf s'} \in G$ is given by
\[
C({\bf s, s'}) = g\left( ||\xi({\bf s}) - \xi({\bf s'})|| \right),
\]
for an arbitrary isotropic covariance function $g$ which is valid on $\mathcal{R}^d$ for $d \geq 1$. In the original paper, \cite{sampGutt} use a two step non-parametric approach to estimation: first, they use multidimensional scaling (see, e.g., \citealp{mardia79}) to generate a two-dimensional coordinate representation (in $D$) of the observation locations in $G$, with inter point distances in $D$ representing sample spatial dispersions. Second, a thin-plate spline interpolation is used to fill in the mapping for all points in the geographic region of interest. 

The original deformation model introduced by \cite{sampGutt} suffered from being unable to quantify the uncertainty introduced in estimating the mapping from $G$ to $D$, so several Bayesian alternatives were subsequently proposed. Two alternatives were suggested independently, due to \cite{damian2001} and \cite{schmidt03}, differing primarily in their specification of a prior distribution on the mapping $\xi(\cdot)$. A major problem in the Sampson and Guttorp paper was that the estimated mapping was often {not} one-to-one and folded over itself, and therefore \cite{damian2001} introduced a prior distribution for the transformed observation locations $(\xi({\bf s}_1), ... , \xi({\bf s}_n))^\top$ which penalizes non-smooth maps, including ones that fold. The parameters for the prior distribution are fixed or calculated from the geographical coordinates. Alternatively, \cite{schmidt03} propose a Gaussian process prior to the mapping $\xi(\cdot)$, again fixing most of the prior parameters or calculating them based on the observation locations. Both of these Bayesian models require intricate MCMC algorithms for model fitting and, due to the high dimensionality of the parameter space, are quite difficult to fit.

All of the deformation methods mentioned thus far require replicates of the spatial data which can, in general, be obtained from detrended spatially-referenced observations over time, although such replications may not always be available. \cite{anderes2008} address this limitation and introduce an approximate likelihood-based deformation approach for a single replicate of densely observed data. In their approach, the transformation $\xi(\cdot)$ is parameterized in terms of local affine transformations, with the parameters of the transformation characterized by an ellipse. The parameters of these ellipses are estimated at each observation location and then smoothed over the spatial region. The likelihood-based approach used here is desirable in that it imposes no requirement on the configuration of observation locations and gives estimates which are easy to obtain and efficient.

}

\subsection{Basis function expansions}

Basis function expansions provide a constructive way to model the nonstationarity in a spatial process. The main idea of basis function expansions comes from the Karhunen-Lo\`{e}ve decomposition of a (mean-zero) spatial process,
\begin{equation} \label{KarLoev}
Y({\bf s}) = \sum_{l=1}^\infty \sqrt{\lambda_l} \hskip0.35ex W_l \hskip0.35ex E_l({\bf s}),
\end{equation}
where the $W_l$ are uncorrelated, standardized random variables, the $\lambda_l$ are eigenvalues, and the $E_l(\cdot)$ are orthogonal eigenfunctions. Choosing the $W_l$ to be Gaussian specifies $Y(\cdot)$ to be a GP (e.g., \citealp{nychka2002}). The $E_l(\cdot)$ being orthogonal eigenfunctions requires 
\[
\int_G E_j({\bf s})E_k({\bf s}) d{\bf s} = \left\{ \begin{array}{cl} 1 & \text{ if } j = k, \\ 0 & \text{otherwise} \end{array} \right.
\]
for all $j, k$. The covariance function of the process (\ref{KarLoev}) is
\begin{equation} \label{KLcov}
C({\bf s, s'}) = \sum_{l=1}^\infty \lambda_l E_l({\bf s})E_l({\bf s'}),
\end{equation}
where the $\lambda_l$ and $E_l(\cdot)$ come from the Fredholm integral equation $
%\begin{equation} \label{fredholm}
\int_G C({\bf s, s'}) E_l({\bf s}) d{\bf s} =  \lambda_l E_l({\bf s'}).
%\end{equation}
$ If the infinite series (\ref{KarLoev}) and (\ref{KLcov}) are truncated to the leading $L$ terms, the finite sum approximation
\begin{equation} \label{KLapprox}
\widehat{C}({\bf s, s'}) = \sum_{l=1}^L \lambda_l E_l({\bf s})E_l({\bf s'})
\end{equation}
is used instead. It can be shown that this truncation minimizes the variance of the truncation error for all sets of $L$ basis functions when the $E_l(\cdot)$ are the exact solutions to the Fredholm equation (\citealp{wikleChapter}) and, as a low-rank representation of the process ${ Y(\cdot)}$, can facilitate computation.  Estimating the covariance function in this way clearly results in a nonstationary covariance structure (i.e., $\widehat{C}({\bf s, s'}) \neq \widehat{C}({\bf s - s'})$). 

The main task with basis function expansions is clearly to model the eigenvalue-eigenfunction pairs  $\{ \lambda_l, E_l(\cdot) \}$. In practice, if an empirical covariance matrix $\widehat{\bf \Sigma}$ can be calculated, the pairs $\{ \lambda_l, E_l(\cdot) \}$ can be approximated by the sample quantities obtained from the spectral decomposition ${\bf \widehat{\Sigma} = \widehat{E} \widehat{D} \widehat{E}^\top}$. Here, ${\bf \widehat{D}} = diag(\widehat{\lambda}_1, \dots , \widehat{\lambda}_n)$, and the columns of $\widehat{\bf E}$ are the estimated eigenvectors $\widehat{E}_l(\cdot)$, which are called empirical orthogonal functions (EOFs). This is the approach taken by \cite{holland}, who use a slight variation of (\ref{KLapprox}) and propose a new covariance function as the sum of a stationary covariance function (including a nugget) and a nonstationary component of the form of (\ref{KLapprox}). The EOFs are calculated from a ``detrended'' empirical covariance matrix which removes the effect of the stationary component; only a single replicate of data was needed for the empirical estimate. \cite{holland} compare this nonstationary covariance function to both a standard isotropic model as well as an exponential covariance function with spatially-varying marginal variances; the nonstationary model greatly reduced the mean square prediction error. %Furthermore, this model has appeal in that it separates the stationary and nonstationary properties of the spatial process.

Alternatively, \cite{nychka2002} use non-orthogonal multiresolution wavelet basis functions in place of the eigenfunction bases in (\ref{KLapprox}), which relaxes the condition that the random variables $\{W_l\}$ are uncorrelated. Multiresolution bases, which have differing ranges of dependence over space, are useful for modeling nonstationary processes because the stochastic properties of the process can be controlled locally while still giving a globally nonstationary covariance function. Computational feasibility is obtained by restricting these basis functions to be translations and scalings of a few fixed functions, and the authors demonstrate the flexibility of the multiresolution model with simulations in which the wavelets approximate standard covariance models very well. The original method in this paper requires observations on a grid;  \cite{matsuo2011} extend this approach to irregularly spaced observations and introduce an EM algorithm to estimate the covariance parameters.

\subsection{Gaussian Markov random field methods}

While not explicitly a Gaussian process model, the SPDE approach of \cite{Lindgren2011} introduces a model for a Gaussian Markov random field (GMRF) which approximates a particular Gaussian process model. GMRFs are popular for areal data, where it is easy to establish the necessary neighborhood structure, and the Markov properties of the model allow major computational gains due to working with the precision matrix instead of the covariance matrix. Unfortunately, GMRFs have (until recently) been poorly suited for spatial models for point-referenced data, since it is difficult to construct a GRMF with a specific spatial correlation structure; on the other hand, such a constructive formulation is natural for GPs. \cite{Lindgren2011} overcome this problem by providing an explicit strategy for constructing a GMRF which corresponds to a GP with a known Mat\'{e}rn covariance function; the link between GMRFs and GPs is given by finding a finite basis function representation which is the solution to a particular SPDE. Nonstationarity is accomplished in this work by allowing the Mat\'{e}rn covariance function parameters (range and marginal variance) to vary over space; \cite{Lindgren2011} suggest using a low-dimensional representation in which these parameters vary smoothly over space according to a log-linear function.

\section{Process convolution or kernel smoothing methods} \label{kernConv}

\subsection{Main result}

Like basis function expansions, the process convolution (also called kernel smoothing or moving average) method is popular because it provides for a constructive approach to specifying a spatial GP. In general, a spatial stochastic process $Y(\cdot)$ on $G \subset \mathcal{R}^d$ can be defined by the kernel convolution 
\begin{equation} \label{kernelconvolution}
Y({\bf s}) = \int_{G} K({\bf s-u}) dW({\bf u})
\end{equation}
(\citealp{thiebaux76}; \citealp{thiebaux_pedder}), where $W(\cdot)$ is a $d$-dimensional stochastic process and $K(\cdot)$ is a kernel function. \cite{higdon2} summarizes the extremely flexible class of spatial statistical models defined using (\ref{kernelconvolution}): see, for example, \cite{Barry1996}, \cite{VerHoef2004}, \cite{Wolpert1999}, \cite{Higdon98}, \cite{VerHoef2004}, and \cite{Barry_VerHoef1998}. The popularity of this approach is due largely to the fact that it is much easier to specify (possibly parametric) kernel functions than a covariance function directly, since the kernel functions only require
\begin{equation*}
\int_{\mathcal{R}^d} K({\bf u}) d{\bf u} <\infty
\end{equation*}
and
\begin{equation*}
\int_{\mathcal{R}^d} K^2({\bf u}) d{\bf u} <\infty,
\end{equation*}
while a covariance function must be a positive definite function. The process $Y(\cdot)$ in (\ref{kernelconvolution}) is a Gaussian process when $W(\cdot)$ is chosen to be Brownian motion or another Gaussian process. If $W(\cdot)$ is specified as Brownian motion, % (e.g., Sections \ref{DPCM_Higdon} and \ref{SVP}), 
an equivalent representation of (\ref{kernelconvolution}) can be obtained by replacing $W(\cdot)$ with a mean-zero process $V(\cdot)$ which has independent increments with variance proportional to the volume of the increment (\citealp{CalderCressie}). That is, the stochastic integral in (\ref{kernelconvolution}) can be re-written as
\begin{equation} \label{kernelconvolution2}
Y({\bf s}) = \int_{G} K({\bf s-u}) V({\bf u})d{\bf u}.
\end{equation}
If $V(\cdot)$ is chosen to be Gaussian white noise (see, e.g., Section \ref{SVP}), then the resulting covariance function is
\[
C({\bf s}, {\bf s}') \equiv E[Y({\bf s}) Y({\bf s}')] = \int_{\mathcal{R}^d} \int_{\mathcal{R}^d} K({\bf s- u}) K({\bf s' - t}) E \left[ V({\bf u}) V({\bf t}) \right] d{\bf u} d{\bf t}.
\]
Because $E \left[ V({\bf u}) V({\bf t}) \right] = 0$ for all ${\bf u} \neq {\bf t}$ and, without loss of generality, $E \left[ V^2({\bf u}) \right] = 1$, the above becomes
\begin{equation} \label{Hig_S}
%\begin{array} {rcl}
C({\bf s}, {\bf s'}) = \int_{\mathcal{R}^d} K({\bf s - u}) K({\bf s' - u}) E \left[ V^2({\bf u}) \right] d{\bf u} = \int_{\mathcal{R}^d}  K({\bf s - u}) K({\bf s' - u}) d{\bf u}.
%\end{array}
\end{equation}
When the kernel function $K(\cdot)$ is constant, (\ref{Hig_S}) can be rewritten using the linear transformation ${\bf t = u-s}$ (with Jacobian 1) as 
\begin{equation} \label{Hig_S2}
C({\bf s}, {\bf s'}) = \int_{\mathcal{R}^d}  K({\bf t}) K({\bf [s' - s] - t}) d{\bf t},
\end{equation}
which is a stationary covariance function.

Three approaches that use the representation in (\ref{kernelconvolution}) to arrive at a nonstationary covariance function are particularly relevant to this dissertation and will be described in greater detail. Two of these methods are based on (\ref{kernelconvolution}) directly; the third method is motivated by (\ref{kernelconvolution}) but relies on the notion of spatially-varying parameters instead of kernel convolution.

\subsection{Discrete process convolution model} \label{DPCM_Higdon}

The first approach, due to \cite{Higdon98}, obtains a nonstationary process in (\ref{kernelconvolution}) by choosing $W(\cdot)$ to be Brownian motion and allowing the kernel to depend on the spatial location, denoted $K_{\bf s}(\cdot)$. In this case, (\ref{kernelconvolution}) can be written as (\ref{kernelconvolution2}) with $V(\cdot)$ as Gaussian white noise; (\ref{kernelconvolution2}) is then approximated by restricting the process to a finite grid of locations $\{{\bf u}_l : l = 1, ... , L\}$ in $G$, i.e., $V({\bf u}_l) \text{ iid } \mathcal{N}(0, \sigma^2_u)$, which defines the approximate (nonstationary) Gaussian process
\begin{equation} \label{disc_spatProc}
\widehat{Y}({\bf s}) = \sum_{l=1}^L K_\bfs({\bf s - u}_l) V({\bf u}_l).
\end{equation}
\cite{Higdon98} further specifies the kernel functions $K_\bfs(\cdot)$ to be $d$-variate Gaussian density functions centered at $\bfs$ with (kernel) covariance matrix $\bfSigma(\bfs)$ (as will be done later; see (\ref{GaussianKernel})). To reduce the computational burden of estimating the spatial convolution kernels $K_\bfs(\cdot)$ globally, \cite{Higdon98} instead uses a local estimation procedure, in which the kernels are estimated for fixed regions over the spatial domain. The convolution kernel for an arbitrary location $\bfs \in G$ is calculated as the weighted average of the locally estimated convolution kernels or ``basis'' kernels. Specifically, following \cite{Higdon98}, specify $\{ {\bf b}_m : m = 1, \dots, M \}$ to be a (coarse) basis grid of evenly spaced locations over $G$ (Higdon chose $M=8$), and define $K_m(\cdot)$ to be the basis kernel centered at each ${\bf b}_m$. Then,
\begin{equation} \label{weightedKernels}
K_\bfs({\bf h}) = \sum_{m=1}^M w_m(\bfs) K_m({\bf h}),
\end{equation}
where the weights are calculated based on distance as
\[
w_m(\bfs) \propto \exp\left\{ -|| \bfs - {\bf b}_m ||^2/2 \right\},
\]
such that $\sum_m w_m(\bfs) = 1$. The Gaussian basis kernels $\{ K_m(\cdot): m = 1,\dots,M\}$ are estimated by first fitting an anisotropic Gaussian variogram model to all of the data within a particular radius of the basis centroids $\{ {\bf b}_m \}$, and then transforming the the estimated variogram parameters back to the parameters of a Gaussian smoothing kernel. The transformation is trivial, since a spatial process specified by a Gaussian variogram is equivalent to a process obtained by convolving a white noise process with a Gaussian kernel in (\ref{kernelconvolution2}) (\citealp{Higdon98}). Using this approach, note that after the variogram parameters (equivalently, the basis kernels) are estimated they are considered fixed for the remainder of the analysis.

\begin{figure}[!t]
\begin{center}
\includegraphics[trim={10 10 10 10mm}, clip, width=\textwidth]{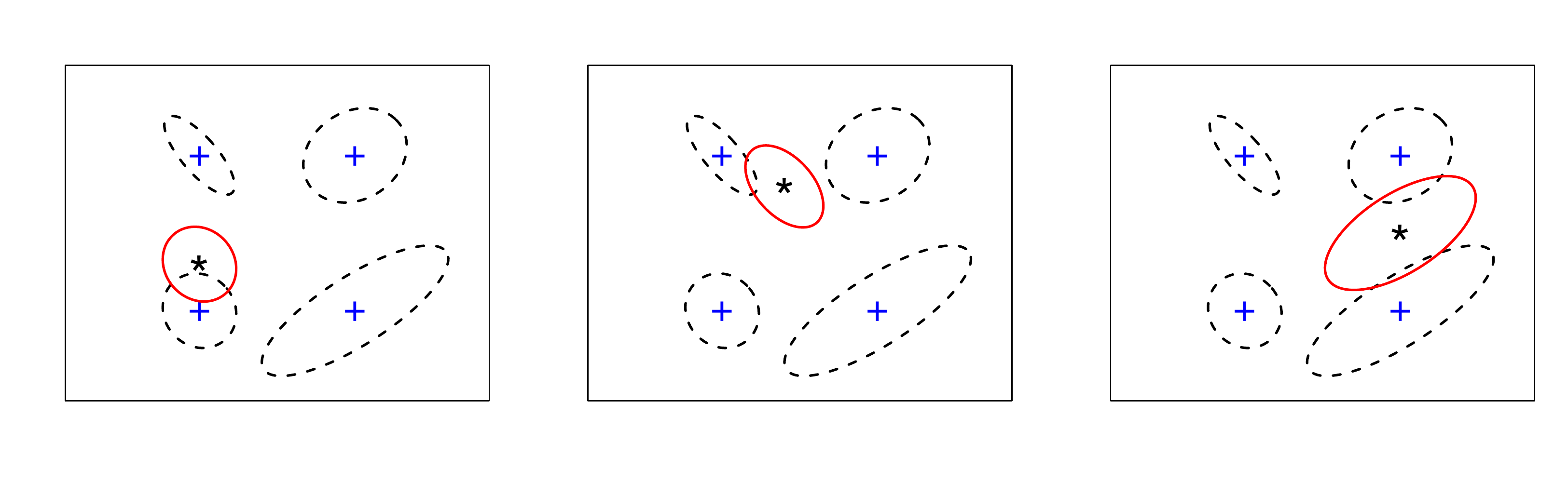}
\caption{Visualization of the discrete basis kernel approach of \cite{Higdon98}. The blue ``$+$'' symbols represent basis locations with fixed (locally estimated) basis kernel functions corresponding to the dashed black ellipses. The ``$\star$'' symbols represent three arbitrary spatial locations and the corresponding ellipses (kernel functions), calculated as in (\ref{weightedKernels}).}
\label{wgtEllp}
\end{center}
\end{figure}

As an illustration of (\ref{weightedKernels}), consider Figure \ref{wgtEllp}. A helpful way to visualize a bivariate Gaussian smoothing kernel function with $2\times 2$ kernel covariance matrix $\bfSigma(\cdot)$ is by plotting the corresponding one standard deviation ellipse of $\bfSigma(\cdot)$. The size and orientation of this ellipse indicate directions in which the range of smoothing is longer or shorter; equivalently, the ellipse represents the local (and spatially-varying) anisotropy of the process, such that directions which have longer ranges of smoothing also have longer ranges of spatial dependence, and vice versa. Figure \ref{wgtEllp} contains a toy example, in which the basis grid is of size $M=4$. The basis locations $\{ {\bf b}_m : m= 1, \dots, 4 \}$ are plotted as blue ``$+$'' symbols. The black dashed ellipse centered at each basis location is the ellipse corresponding to the locally estimated anisotropic Gaussian variogram parameters (the basis kernel functions), which are considered fixed. The black ``$\star$'' symbols represent three arbitrary spatial locations $\{ \bfs_1, \bfs_2, \bfs_3 \}$, and the red ellipse centered at each of these locations corresponds to the covariance matrix of the kernel functions $K_{\bfs_1}(\cdot)$, $K_{\bfs_2}(\cdot)$, $K_{\bfs_3}(\cdot)$, which are a weighted average of the dashed black ellipses following (\ref{weightedKernels}).

\subsection{Convolution of locally stationary processes} \label{fuentesApproach}

An alternative approach to specifying a nonstationary process model using (\ref{kernelconvolution}) is due to \cite{fuentes02}, who chooses the kernel function to be constant over $G$ while allowing the spatial stochastic process to vary over the spatial region. 
Previously, $W(\cdot)$ was simply a general $d$-dimensional stochastic process; instead of a single stochastic process, \cite{fuentes02} replaces $W(\cdot)$ with a parametric family of spatially-varying stationary Gaussian processes, denoted $\{ \widetilde{Y}_{\bf s}(\cdot) : \bfs \in G \}$, such that the processes are orthogonal, i.e.,
\[
\widetilde{Y}_{\bf s}(\cdot) \perp \widetilde{Y}_{\bfs'}(\cdot) \text{ for } \bfs \neq \bfs'.
\]
%Making this substitution, (\ref{kernelconvolution}) becomes
%\[
%Y({\bf s}) = \int_{G} K({\bf s-u}) d\widetilde{Y}_{\bf u}({\bf u}).
%\]
Each $\widetilde{Y}_\bfs(\cdot)$ can be thought of arising from (\ref{kernelconvolution2}) (\citealp{fuentes2002b}), so that $\widetilde{Y}_\bfs(\cdot)$ has the stationary covariance function derived in (\ref{Hig_S2}). Alternatively, $\widetilde{Y}_\bfs(\cdot)$ could simply be specified as having one of the parametric covariance functions outlined in Section \ref{statCovModels} such that the parameter vector depends on location, i.e., $\bftheta(\bfs)$ (\citealp{fuentes02}). Regardless of how the processes $\{ \widetilde{Y}_{\bf s}(\cdot) : \bfs \in G \}$ are defined, let $\widetilde{C}_{\bf s}(\cdot)$ denote the corresponding stationary covariance function assigned to each process.

The intuition behind this approach is that a nonstationary spatial process $Y(\cdot)$ is constructed by convolving the locally stationary processes $\{ \widetilde{Y}_{\bf s}(\cdot) : \bfs \in G \}$. In this case, the resulting (nonstationary) covariance function of $Y(\cdot)$ is
\begin{equation} \label{Fuen_NS}
C({\bf s}, {\bf s'}) = \int_{G}  K({\bf s - u}) K({\bf s' - u}) \widetilde{C}_{\bf u} ( {\bf s - s'} ) d{\bf u}.
\end{equation}
In practice, the integral in (\ref{Fuen_NS}) necessitates an approximation similar to the one in (\ref{disc_spatProc}), resulting in the approximated covariance function
\begin{equation} \label{disc_Fuen_NS}
\widehat{C}({\bf s}, {\bf s'}) = \sum_{l=1}^L  K({\bf s - u}_l) K({\bf s' - u}_l) \widetilde{C}_{{\bf u}_l} ( {\bf s - s'} ) d{\bf u},
\end{equation}
used in \cite{fuentes2001}, \cite{fuentes2002b}, and \cite{fuentes02}. This finite-sum representation only approximates the true covariance function implied by (\ref{Fuen_NS}), although \cite{fuentes02} shows that the spatial process $\widehat{Y}(\cdot)$ corresponding to the approximated covariance in (\ref{disc_Fuen_NS}) converges in distribution to the true process $Y(\cdot)$ corresponding to the true covariance (\ref{Fuen_NS}) under infill asymptotics.

\subsection{Spatially-varying parameters} \label{SVP}

The final method described here which uses (\ref{kernelconvolution2}) to construct a nonstationary spatial process model is due to \cite{Higdon99}. Again specifying the process $V(\cdot)$ to be Gaussian white noise, the resulting covariance function was derived in (\ref{Hig_S}). If the kernel functions are spatially-varying, i.e., $K_{\bf s}(\cdot)$, then (\ref{Hig_S}) becomes
\begin{equation} \label{Hig_NS}
C({\bf s}, {\bf s'}) = \int_{\mathcal{R}^d} K_{\bf s}({\bf u}) K_{\bfs'}({\bf u}) E \left[ V^2({\bf u}) \right] d{\bf u}  = \int_{\mathcal{R}^d}  K_{\bf s}({\bf u}) K_{\bf s'}({\bf u}) d{\bf u},
\end{equation}
which is a nonstationary covariance function. In general, any kernel function can be used for (\ref{Hig_NS}), although most choices prevent explicit calculation of the integral in (\ref{Hig_NS}). However, if the $K_{\bf s}(\cdot)$ are $d$-variate Gaussian densities centered at $\bfs$ with covariance matrix $\bfSigma(\bfs)$ (as in Section \ref{DPCM_Higdon}), the integral in (\ref{Hig_NS}) can be calculated analytically. That is, if 
\begin{equation} \label{GaussianKernel}
K_{\bf s}({\bf u}) = \frac{1}{(2\pi)^{d/2} |\bfSigma(\bfs)|^{1/2}} \exp \left\{ -\frac{1}{2} ({\bf s - u})^\top \bfSigma(\bfs)^{-1} ({\bf s - u}) \right\},
\end{equation}
then we have the following:
\begin{equation} \label{GaussianKernConv}
\begin{array} {rcl}
C^H({\bf s}, {\bf s}') & =  &\int_{\mathcal{R}^d} (2\pi)^{-d/2} |\bfSigma(\bfs)|^{-1/2} \exp \left\{ -\frac{1}{2} ({\bf s - u})^\top \bfSigma(\bfs)^{-1} ({\bf s - u}) \right\} \\[2ex]
 & & \hskip3ex \cdot \hskip1ex (2\pi)^{-d/2} |\bfSigma(\bfs')|^{-1/2} \exp \left\{ -\frac{1}{2} ({\bf s' - u})^\top \bfSigma(\bfs')^{-1} ({\bf s' - u}) \right\} d{\bf u}.
\end{array}
\end{equation}
The ``$H$'' superscript is used throughout to denote the covariance function used in \cite{Higdon99}. Defining $\phi_{Y, q}(\cdot)$ to be the density function for a $q$-variate Gaussian random variable $Y$, (\ref{GaussianKernConv}) can be represented as
\begin{equation} \label{convolution}
C^H({\bf s}, {\bf s}') = \int_{\mathcal{R}^d} \phi_{F, d} ({\bf u - s}) \cdot \phi_{G, d} ({\bf u}) d{\bf u} = \int_{\mathcal{R}^d} \phi_{[F,G], 2d}({\bf u - s, u}) d{\bf u},
\end{equation}
where $F \sim N_d({\bf 0}, \bfSigma(\bfs))$ and $G \sim N_d({\bf s}', \bfSigma(\bfs'))$, such that $F$ and $G$ are independent. Define the linear change of variable (with Jacobian 1) $A=G-F$ and $B=G$, so that if $G=u$ and $F=u-s$ then $a=s$ and $b=u$. Then, equation (\ref{convolution}) becomes
\[
\int_{\mathcal{R}^d} \phi_{(A,B), 2d}({\bf s, u}) d{\bf u} = \phi_{A, d}({\bf s}),
\]
where from independence of $F$ and $G$ we are able to derive that $A \sim N_d({\bf s}, \bfSigma(\bfs) + \bfSigma(\bfs'))$. Thus, combining the above, 
\begin{equation} \label{cov22}
\begin{array} {rcl}
C^H({\bf s}, {\bf s}') & = & \int_{\mathcal{R}^d}  K_{\bf s}({\bf u}) K_{\bf s'}({\bf u}) d{\bf u} \\[2ex]
 & = & \phi_{A, d}({\bf s}) \\[2ex]
 & = & (2\pi)^{-d/2} |\bfSigma(\bfs) + \bfSigma(\bfs')|^{-1/2} \exp\left\{-\frac{1}{2} (\bfs - \bfs')^\top [\bfSigma(\bfs) + \bfSigma(\bfs')]^{-1}(\bfs - \bfs') \right\} \\[2ex]
  & = & \left(2\sqrt{\pi}\right)^{-d}  \left|\frac{{\bf \Sigma}(\bfs) + \bfSigma(\bfs')}{2}\right|^{-1/2} \rho\left(\sqrt{Q(\bfs, \bfs')}\right),
\end{array}
\end{equation}
where again $\bfSigma(\bfs)$ is the $d \times d$ covariance matrix for the Gaussian kernel function centered at location ${\bf s}$, 
\begin{equation} \label{Q}
Q({\bf s, s'}) = {\bf (s-s')}^\top \left(\frac{{\bf \Sigma(s) + \Sigma(s')}}{2}\right)^{-1}{\bf (s-s')}
\end{equation}
is a scaled squared distance, and $\rho(h) = \exp\{-h^2\}$ is the standard Gaussian correlation function (i.e., $\mathcal{M}_{\infty}(\cdot)/\sigma^2$ with $\phi=1$). To avoid confusion, we will henceforth refer to $\bfSigma(\bfs)$ as the ``kernel matrix'' for location $\bfs$ (following \citealp{PacScher}). 

\cite{Higdon99} use (\ref{cov22}) directly, which boils down to specifying a model for the kernel matrices $\bfSigma(\bfs)$. For $d=2$, when the kernel functions are bivariate Gaussian densities, \cite{Higdon99} note that there is a one-to-one mapping between a bivariate mean-zero Gaussian density and its one standard deviation ellipse, so the spatially-varying kernel matrices $\bfSigma(\bfs)$ are modeled using a spatially-varying family of ellipses. Specifically, \cite{Higdon99} parameterizes the ellipses to be spatially-varying by allowing the coordinates of the focal points of $\bfSigma(\bfs)$ to be spatially-varying; these coordinates are assigned stationary Gaussian process prior distributions.

However, as discussed in \cite{PacScher}, using Gaussian kernel functions (equivalently, a Gaussian correlation function) as in (\ref{cov22}) has the undesirable property of giving process realizations which are infinitely differentiable and therefore too smooth for most environmental applications (see also \citealp{SteinBook}). As an intuitive alternative, \cite{PacScher} suggest substituting any valid isotropic correlation function $g(\cdot)$ in place of $\rho(\cdot)$ of (\ref{cov22}), that is,
\begin{equation} \label{CPS}
C^{PS}({\bf s}, {\bf s}') = \left(2\sqrt{\pi}\right)^{-d}  \left|\frac{{\bf \Sigma}(\bfs) + \bfSigma(\bfs')}{2}\right|^{-\frac{1}{2}} g\left(\sqrt{Q(\bfs, \bfs')}\right).
\end{equation}
(The ``$PS$'' superscript is used to denote the contribution of \citealp{PacScher}). However, it must be proven that $C^{PS}$ is a valid covariance function, because unfortunately $\sqrt{Q({\bf s, s'})}$ is not a distance metric. Following \cite{PacScher}, proving this fact is an application of Theorem 2 of \cite{schoenberg} (p. 817), which states that  the class of functions that are non-negative definite on the Hilbert space is identical to the class of functions of the form
$$
R(h) = \int_0^\infty \exp\{-h^2t\} dH(t),
$$
where $H(\cdot)$ is nondecreasing and bounded and $t\geq 0$. This class of functions is identical to the class of functions non-negative definite on $\mathcal{R}^d, d\geq 1$ (\citealp{schoenberg}), which is also identical to the class of valid isotropic correlation functions on $\mathcal{R}^d, d\geq 1$ (from Bochner's Theorem; see Section \ref{2ndOrderProps}). Using this result, we can rewrite (\ref{cov22}) as

\begin{tabular}{r}
\noindent $C^{PS}({\bf s}, {\bf s}') = \left(2\sqrt{\pi}\right)^{-d}  \left|\frac{{\bf \Sigma}(\bfs) + \bfSigma(\bfs')}{2}\right|^{-\frac{1}{2}} R\left(\sqrt{Q(\bfs, \bfs')}\right)$
\end{tabular}
\begin{equation*} 
\begin{array} {rcl}
\hskip6ex & = & \int_0^\infty  \left(2\sqrt{\pi}\right)^{-d}  \left|\frac{{\bf \Sigma}(\bfs) + \bfSigma(\bfs')}{2}\right|^{-\frac{1}{2}}  \exp\{-Q(\bfs, \bfs') \cdot t\} \>\> dH(t) \\[2ex]
 
\hskip6ex & = & \int_0^\infty \left(2\sqrt{\pi}\right)^{-d}  \left|\frac{{\bf \Sigma}(\bfs) + \bfSigma(\bfs')}{2}\right|^{-\frac{1}{2}}   \exp\left\{- (\bfs -\bfs')^\top \hskip-0.75ex \left( \frac{ \bfSigma(\bfs)+ \bfSigma(\bfs')}{2t} \right)^{-1} \hskip-0.75ex (\bfs - \bfs') \right\}  dH(t) \\[2ex]
 
\hskip6ex & = & \int_0^\infty t^{-d/2} \Big[ \int_{\mathcal{R}^d}  K^t_{\bfs}({\bf u}) K^t_{\bfs'}({\bf u}) d{\bf u}  \Big] dH(t),
\end{array}
\end{equation*}
where $K^t_{\bfs}({\bf u})$ is a Gaussian density centered at ${\bfs}$ with (kernel) covariance matrix ${\bfSigma}^t(\bfs) = {\bfSigma}(\bfs)/t$. The final equality comes from reversing the algebra in (\ref{cov22}).

Using this representation, we can next check to ensure that this function is non-negative definite, from (\ref{PD}). Since we are dealing with complex numbers, write each $a_j = \alpha_j + i \beta_j$, where $\alpha_j$ and $\beta_j$ are real numbers. Then, 

\begin{tabular}{r}
\noindent $\sum_{k=1}^n \sum_{j=1}^n (\alpha_k + i \beta_k) (\alpha_j - i \beta_j) C^{PS}({\bf s}_k, {\bf s}_j)$
\end{tabular}
\begin{equation} \label{PDproof} 
\begin{array} {rcl}
 & = & \sum_{k=1}^n \sum_{j=1}^n \alpha_k \alpha_j C^{PS}({\bf s}_k, {\bf s}_j) - \sum_{k=1}^n \sum_{j=1}^n i\alpha_k \beta_j C^{PS}({\bf s}_k, {\bf s}_j) \\[2ex]
 &    & + \sum_{k=1}^n \sum_{j=1}^n i\alpha_j \beta_k C^{PS}({\bf s}_k, {\bf s}_j) - \sum_{k=1}^n \sum_{j=1}^n i^2 \beta_j \beta_k C^{PS}({\bf s}_k, {\bf s}_j) \\[2ex]
 & = & \sum_{k=1}^n \sum_{j=1}^n \alpha_k \alpha_j C^{PS}({\bf s}_k, {\bf s}_j) + \sum_{k=1}^n \sum_{j=1}^n \beta_j \beta_k C^{PS}({\bf s}_k, {\bf s}_j).
\end{array}
\end{equation}
The last equality follows because $C^{PS}$ is symmetric. Because both the $\{\alpha_j\}$ and $\{\beta_j\}$ are real numbers, each part of (\ref{PDproof}) becomes 

\begin{tabular}{r}
\noindent $\sum_{k=1}^n \sum_{j=1}^n \alpha_k \alpha_j C^{PS}({\bf s}_k, {\bf s}_j)$
\end{tabular}
\begin{equation} \label{PDproof2} 
\begin{array} {rcl}
 & = & \sum_{k=1}^n \sum_{j=1}^n \alpha_k \alpha_j \int_0^\infty t^{-d/2} \Big[ \int_{\mathcal{R}^d}  K^t_{\bfs_k}({\bf u}) K^t_{\bfs_j}({\bf u}) d{\bf u}  \Big] dH(t) \\[2ex]
 
 & = & \int_0^\infty  \int_{\mathcal{R}^d} t^{-d/2}  \Big(\sum_{k=1}^n  \alpha_k K^t_{\bfs_k}({\bf u}) \Big) \Big(\sum_{j=1}^n \alpha_j K^t_{\bfs_j}({\bf u})\Big) d{\bf u} \> dH(t) \\[2ex]
 
 & = & \int_0^\infty  \int_{\mathcal{R}^d} t^{-d/2}  \Big(\sum_{k=1}^n  \alpha_k K^t_{\bfs_k}({\bf u}) \Big)^2 d{\bf u} \> dH(t) \\[2ex]
 
 & \geq & 0.
\end{array}
\end{equation}
The same argument applies to the second summation in (\ref{PDproof}). Therefore, the entire quantity is non-negative, and $C^{PS}$ is non-negative definite (and hence a valid covariance function), as required.

Building off the ideas in \cite{paciorek2003}, \cite{stein2005} proves that a slight generalization of (\ref{cov22}) still gives a valid covariance function. Specifically,
\begin{equation} \label{NScov}
C^{NS}({\bf s}, {\bf s'}; \boldsymbol{\theta}) = \sigma({\bf s}) \sigma({\bf s'}) \frac{\left|{\bf \Sigma(s)}\right|^{1/4} \left|{\bf \Sigma(s')}\right|^{1/4} }{ \left|\frac{{\bf \Sigma(s) + \Sigma(s')}}{2}\right|^{1/2} } g \left( \sqrt{Q({\bf s, s'})} \right),
\end{equation}
is a valid, nonstationary, parametric covariance function for $\mathcal{R}^d, d \geq 1$, where $g(\cdot)$ is again chosen to be a valid correlation function for $\mathcal{R}^d, d \geq 1$. In (\ref{NScov}), $\boldsymbol{\theta}$ is a generic parameter vector of all variance and covariance parameters, $\sigma(\cdot)$ represents a spatially-varying standard deviation, and the kernel matrix ${\bf \Sigma}(\cdot)$ again represents the spatially-varying local anisotropy (controlling both the range and direction of dependence). The primary contribution of this result is that if $g(\cdot)$ is chosen to be the Mat\'{e}rn correlation function, the smoothness can also be spatially-varying, in which case the smoothness for (\ref{NScov}) is $[\kappa({\bf s}) + \kappa({\bf s}')]/2$ (\citealp{stein2005}). However, another reason for re-writing (\ref{CPS}) in this way is that the variance of $Y(\cdot)$ becomes
\[
\text{Var}[Y(\bfs)] \equiv C^{NS}({\bf s}, {\bf s}; \boldsymbol{\theta}) = \sigma^2({\bf s}),
\]
which does not depend on $\bfSigma(\cdot)$. While using (\ref{NScov}) no longer requires the notion of kernel convolution, we still refer to ${\bf \Sigma}(\cdot)$ as the kernel matrix, since it was originally defined in terms of the kernel functions. 

The covariance function (\ref{NScov}) is extremely flexible, and has been used in various forms throughout the literature. \cite{PacScher} fix $\kappa({\bf s}) \equiv \kappa$ and $\sigma({\bf s}) \equiv \sigma$ ($\kappa$ and $\sigma$ unknown constants) for all ${\bf s}$; the kernel matrices are again modeled by assigning the components of their spectral decompositions to have independent, stationary Gaussian process priors. \cite{Anderes_Stein} find that it is difficult to separate the effect of ${\bf \Sigma(s)}$ and $\kappa({\bf s})$ if (\ref{NScov}) is used directly; instead, they constrain the kernel matrices to be a multiple of the identity matrix and introduce a separate model for $\kappa(\cdot)$. \cite{kleiber2012} use (\ref{NScov}) directly for multivariate spatial processes, and \cite{Katzfuss2013} models the spatially-varying parametric quantities using basis function approximations.

\section{Methods for including covariate information in a covariance function} \label{cov_in_cov}
%\subsection{Including covariate information}
Building off the intuition of mean regression, more recent methodology uses the idea that covariate information might play a useful role in specifying the covariance structure of a spatial process. By covariate information we mean spatially-varying, observable quantities which can either be collected at all prediction locations of interest or in some way interpolated from nearby observations (for example, elevation, wind speed or direction, soil quality, proximity to a pollution source or geographical feature, etc.). In general, the argument for using covariate information in a covariance function is both interpretational and computational. First, the major drawback to not using covariate information to model spatial dependence is that it becomes difficult to understand why the process exhibits nonstationary behavior, i.e., how the dependence structure changes over space. Introducing covariates in a covariance function is a natural way to impose the desired second-order nonstationarity and allows for an explanation of the spatially-varying dependence structure. Secondly, many of the ``non-covariate'' models are highly parameterized and therefore difficult to fit, since it is hard to estimate nonstationary behavior using only a single realization of a spatial process. As in mean regression, using covariates allows the dimension of the requisite parameter space to be greatly reduced, facilitating computation.

Indeed, some work has been done to expand the aforementioned nonstationary approaches to incorporate covariate information. \cite{schmidt11} introduce covariate information in a deformation model by incorporating covariates in the mapping from geographic space to the deformed space in two ways. Previously, if the dimension of $G$ is $d$, the mapping $\xi(\cdot)$ was from $\mathcal{R}^d \rightarrow \mathcal{R}^d$; the first strategy in this paper involves extending the mapping to be from $\mathcal{R}^d \rightarrow \mathcal{R}^{d+p}$, $p>0$. In general, the additional $p$ dimensions (``axes'') of the deformed space are useful in avoiding the non-bijective property encountered by \cite{sampGutt}, but the extra dimensions can also be used to incorporate covariate information. That is, if it is known that a set of $p$ covariates might prove useful in explaining spatial correlations, then they can be used for the additional coordinates of $\xi({\bf s})$. Unfortunately, if the size of the parameter space was large in the Bayesian deformation methods described previously, the size of the parameter space is now even larger. To address this problem, \cite{schmidt11} also propose a so-called ``projection model,'' which involves fixing the form of the mapping $\xi(\cdot)$ to be affine, calculating distances using a Mahalanobis distance. This simplification greatly eases the implementation of an MCMC algorithm, but since the $D$ space is now simply the $G$ space augmented by useful covariate information, it is not entirely clear how the anisotropy introduced by the Mahalanobis distance in the $D$ space implies nonstationarity in the $G$ space.

Covariate information was first introduced in the \cite{Higdon99} version of a  process convolution model by \cite{calder08}, who proposed a spatio-temporal model which convolved spatially independent autoregressive processes. Following \cite{Higdon99}, Calder chose the kernels to be Gaussian, but used covariate information to fix the kernel function parameters. More specifically, these parameters were estimated using variogram techniques and then considered fixed for the remainder of the otherwise Bayesian analysis. The covariates used in this paper were wind direction and speed; because this covariate is temporally-varying, the kernel parameters were also able to vary with time, naturally resulting in a nonstationary space-time model. Alternatively, \cite{ViannaNeto} introduce a convolution approach that also incorporates directional covariates, and compare a model using the closed-form covariance function of \cite{PacScher} with a discrete sum approximation; both models use directional covariates to model the parameters of the kernel functions. \cite{ViannaNeto} found that the latter, more parsimonious model was far easier to implement using MCMC. 

\cite{reich2011} propose a spatio-temporal model with covariate information for the \cite{fuentes02} kernel-smoothing method, again using a discrete sum approximation (with covariance function as in (\ref{disc_Fuen_NS})). This model accomplishes greater generality than either \cite{calder08} or \cite{ViannaNeto} in the sense that any type of covariate information can be included. The model in this paper is actually somewhat of a hybrid between the \cite{Higdon99} and \cite{fuentes02} models: the $L$ stationary processes with covariance functions $\widetilde{C}_{{\bf u}_l}(\cdot)$ are exponential with different ranges $\{\phi_l, l=1, ... , L\}$, but the kernel functions also vary with the index (i.e., $K({\bf s - u}_l) \equiv K_l({\bf s})$). Specifically, the $L$ kernel functions are chosen to be a log-linear function of generic covariate information at location ${\bf s}$. An interesting discussion is given on the properties of the resulting covariance function, although (like \citealp{ViannaNeto}) there is little information on how the covariate information impacts the covariance function. %A fully Bayesian analysis is conducted via MCMC.

A recent extension of the \cite{Lindgren2011} work is given by \cite{spde13}, who show that using stochastic differential equations for spatial modeling allows covariate information to be easily introduced in the dependence structure. As previously mentioned, \cite{Lindgren2011} suggest using a log-linear function to model spatially-varying Mat\'{e}rn parameters; \cite{spde13} take this idea one step further and impose a log-linear function of covariate information for the spatially-varying parameters. Specifically, elevation is used in a model for annual precipitation data, although their framework allows for any number or type of covariate information to be included. As a GMRF method, the authors suggest integrated nested Laplace approximations (INLA) for inference and estimation.

\section{Discussion}

The broad literature on nonstationary spatial modeling discussed in Sections \ref{ch2LR}, \ref{kernConv}, and \ref{cov_in_cov} provides flexible classes of covariance functions that more appropriately model the covariance in a spatial process. However, most are also highly complex and require intricate model-fitting algorithms, making it very difficult to replicate their results in a general setting. As a result, when new nonstationary methods are developed, their performance is usually compared to stationary models, for which robust software is available. In order to more accurately evaluate new nonstationary methods, pre-packaged and efficient options for fitting existing nonstationary models must be made available.

To address this need, \cite{RisserCalder2016} develop a nonstationary spatial Gaussian process model with (\ref{NScov}) as the covariance function for $Y(\cdot)$. The model is simplified in that the locally-varying geometric anisotropies are modeled using a ``mixture component'' approach, similar to the discrete mixture kernel convolution approach in \cite{Higdon98}, while also allowing the underlying correlation structure to be specified by the modeler. The model is extended to allow other properties to vary over space as well, such as the process variance, nugget effect, and smoothness. An additional degree of efficiency is gained by using local likelihood techniques to estimate the spatially-varying features of the spatial process; then, the locally estimated features are smoothed over space, similar in nature to the approach of \cite{fuentes02}. Furthermore, \cite{RisserCalder2016} also present and describe the {\tt convoSPAT} package for {\tt R} for conducting a full analysis of point-referenced spatial data using a nonstationary spatial Gaussian process model. The primary contribution of the package is to offer efficient model-fitting for nonstationary, point-referenced spatial data, even when the size of the data is relatively large (on the order of $n=1000$). The package is able to handle both a single realization of a spatial process as well as replicates.

Section \ref{cov_in_cov} provides a discussion of various strategies in the literature that use covariate information to account for nonstationary behavior in the second-order properties of a spatial process. While all of these approaches have been shown to be successful, limitations remain. Several methods are only appropriate for directional covariates (e.g., \citealp{ViannaNeto}) while others are not fully Bayesian (e.g., \citealp{calder08}). More seriously, while successfully including covariate information, many of the methods fail to address the issue of characterizing \textit{how} a spatially-varying covariate impacts the covariance function. For example, \cite{reich2011} use covariate information to model the weights for each of the locally stationary spatial processes, not the spatial dependence properties of these processes. 

\cite{Risser15} address these limitations by proposing new methodology which is fully Bayesian and allows covariate information to be included directly in a model for the spatial dependence properties of the resulting covariance function. Furthermore, the model is able to accommodate any type of covariate information, be it scalar or directional, discrete or continuous, and yields a parsimonious parameterization so that a relatively fast, stable, and efficient model fitting algorithm can be implemented. Finally, the parameters allow for interpretations of how the covariate impacts the spatial dependence, and, given parameter estimates, changes in spatial dependence over the region of interest are easily visualized. The resulting model is applicable in any geostatistical setting in which a spatial Gaussian process model is appropriate; e.g., modeling and prediction of environmental, meteorological, and pollution- or disease-related processes. 

In conclusion, in many applications the relevant covariate information for modeling the first- and/or second-order properties of a spatial process is not available for every prediction location (or even observed location) of interest. Therefore, the methodology of \cite{reich2011}, \cite{ViannaNeto}, and \cite{spde13}, and even \cite{Risser15} cannot be used, unless the covariate itself can be somehow smoothed over space (as in \citealp{ViannaNeto}). However, given that the covariate drives the second-order properties, this may not be desirable or even possible in the case of a non-continuous covariate. 

To navigate this limitation, \cite{Risseretal2016} incorporate Bayesian CART for treed covariate segmentation in a manner that allows the first- and second-order properties of the spatial process of interest to be similar in regions where the distribution of a covariate is homogeneous. Using Bayesian model averaging to account for uncertainty in the segmentation process and to accommodate multiple covariate processes, \cite{Risseretal2016} outline an algorithm for generating predictions at unobserved locations based on the covariate-driven nonstationary model. Additionally, the model yields much faster computation relative to traditional spatial models.

\bibliographystyle{imsart-nameyear} \bibliography{MDRResearch}

\end{document}